\documentclass[prl,twocolumn,superscriptaddress]{revtex4-1}

\usepackage{hyperref}
\usepackage{rotating}
\usepackage{graphicx}
\usepackage{latexsym}
\usepackage{amssymb}
\usepackage{amsfonts}
\usepackage{soul}
\usepackage{color}
\usepackage{amsmath}
\usepackage{color}
\usepackage{lipsum}
\usepackage{natbib}

\begin{document}

\title{Multiband material with a quasi-1D band as a robust high-temperature superconductor}

\author{T. T. Saraiva}
\affiliation{National Research University Higher School of Economics, 101000, Moscow, Russia}

\author{P. J. F. Cavalcanti}
\affiliation{Departamento de Fisica, Universidade Federal de Pernambuco, Cidade Universitaria, 50670-901 Recife-PE, Brazil}

\author{A. Vagov}
\affiliation{Institut f\"{u}r Theoretische Physik III, Bayreuth Universit\"{a}t, Bayreuth 95440, Germany}

\author{A. S. Vasenko}
\affiliation{National Research University Higher School of Economics, 101000, Moscow, Russia}
\affiliation{I.E. Tamm Department of Theoretical Physics, P.N. Lebedev Physical Institute, Russian Academy of Sciences, 119991 Moscow, Russia}

\author{A. Perali}
\affiliation{School of Pharmacy, Physics Unit, University of Camerino, I-62032 Camerino, Italy}

\author{L. Dell'Anna}
\affiliation{Dipartimento di Fisica e Astronomia ``Galileo Galilei", Universit\`{a}  di Padova, Via Marzolo 8, 35131 Padova, Italy}

\author{A. A. Shanenko}
\email[]{shanenkoa@gmail.com}
\affiliation{National Research University Higher School of Economics, 101000, Moscow, Russia}
\affiliation{Departamento de Fisica, Universidade Federal de Pernambuco, Cidade Universitaria, 50670-901 Recife-PE, Brazil}

\date{\today}
\begin{abstract}
It is well known that superconductivity in quasi-one-dimensional (Q1D) materials is hindered by large fluctuations of the order parameter. They reduce the critical temperature and can even destroy the superconductivity altogether. Here it is demonstrated that the situation changes dramatically when a Q1D pair condensate is coupled to a higher-dimensional stable one, as in recently discovered multiband Q1D superconductors. The fluctuations are suppressed even by vanishingly small pair-exchange coupling between different band condensates and the superconductor is well described by the mean field theory. In this case the low-dimensionality effects enhance the coherence of the system instead of suppressing it. As a result, the critical temperature of the multiband Q1D superconductor can increase by orders of magnitude when the system is tuned to the Lifshitz transition with the Fermi level close to the edge of the Q1D band.
\end{abstract}

\maketitle

It is a common knowledge that superconductivity in 1D systems is suppressed due to large fluctuations of the order parameter. The superconducting state can still be achieved when several 1D structures (parallel chains of molecules or atoms) are coupled one to another, creating a weakly coupled matrix. Earlier theoretical studies demonstrated that such Q1D materials can superconduct~\cite{Efetov1974,Gorkov1975,Klemm1976,Schulz1983} but the fluctuations are still large, reducing the critical temperature $T_c$ significantly~\cite{Efetov1974}. These predictions were confirmed by the discovery of the low-temperature superconductivity in Bechgaard salts - organic Q1D superconductors~\cite{Bechgaard1980,Lebed2008}.

Subsequent theoretical efforts were focused on finding the conditions under which the critical temperature of the Q1D superconductors could be increased rather than reduced. In particular, it was suggested that such an increase can be achieved in the vicinity of the Lifshitz transition at which the chemical potential approaches the edge of the Q1D single-particle energy band~\cite{Perali1996,Perali1997, Shan2006, Shan2010,Mazz2017}. However, the fluctuations, that are already very large in the presence of the Q1D effects, are additionally enhanced due to the Bose-like character of the pairing which tends to further deplete the condensate. The enhancement of $T_c$ was found for weakly interacting stripes, formed due to a particular transformation of the antiferromagnetic insulator~\cite{Fradkin1998,Fradkin2004}.
The effect requires, however, a subtle balance of different interplaying physical mechanisms, relevant for superconducting cuprates.

Recently the interest in Q1D superconductors has been boosted by the discovery of ${\rm Cr}_2{\rm As}_3$-chain based materials~\cite{Bao2015,Tang2015A,Tang2015B,Wu2019}. Results of the first principle calculations of the electronic band structure of those compounds led to a conclusion that they are multiband systems with some of the contributing bands being Q1D~\cite{Jiang2015,Wu2019} -  multiband Q1D superconductors. For example, ${\rm K}_2{\rm Cr}_3{\rm As}_3$~\cite{Jiang2015} and ${\rm K}{\rm Cr}_3{\rm As}_3{\rm H}_x$~\cite{Wu2019} have two Q1D sheets coexisting with one 3D sheet in the Fermi surface. Furthermore, it was demonstrated that in ${\rm K}{\rm Cr}_3{\rm As}_3{\rm H}_x$ the Fermi level can be shifted by changing the ${\rm H}$-intercalation~\cite{Wu2019}, which gives rise to alterations in topology of the Fermi surface manifested in the Lifshitz transitions.  

In this work we show that the advent of multiband Q1D superconductors opens up a fundamental opportunity to achieve superconductivity at high temperatures. It has already been demonstrated that the presence of the pair-exchange coupling between different bands can reduce the fluctuations due to the multiband screening mechanism~\cite{Perali2000,Wolf2017,Salasnich2019}. Motivated by this result as well as by the recent experimental advances, here we investigate a two-band material with the coupled Q1D and 3D Bardeen-Cooper-Schrieffer (BCS) condensates and demonstrate that under fairly general conditions, it is a robust mean-field superconductor whose critical temperature can be significantly increased by tuning the Lifshitz transition at the edge of the Q1D band.

We assume the $s$-wave pairing in both Q1D and 3D bands with the Josephson-like interband transfer of Cooper pairs. Superconductivity in this system is described by the standard two-band model introduced in Refs.~\cite{Suhl1959,Mos1959}. The intraband and interband pair-exchange couplings are determined by the real matrix $\check{g}$, with the elements $g_{\nu\nu'}=g_{\nu'\nu}$~($\nu=1,2$). For simplicity we consider the parabolic dispersion of the single-particle energy in both bands. For the same reason the Fermi surface of the 3D band ($\nu=1$) is taken spherically symmetric. The principal axis of the Q1D band ($\nu=2$) is chosen parallel to the $z$-axis. In the $x$ and $y$ directions the Q1D energy dispersion is degenerate and we assume the effective finite integral of the density of states (DOS) for both these directions. The band-dependent single-particle energies, shifted by the chemical potential $\mu$, are thus given by
\begin{equation}
\xi_{{\bf k}}^{(1)}=\varepsilon_{0} +\frac{\hbar^2 {\bf k}^2}{2m_1} - \mu, \quad \xi_{{\bf k}}^{(2)}=\frac{\hbar^2 k^2_z}{2m_2} - \mu,\;
\label{disp12}
\end{equation}
where $m_{1,2}$ are the effective band masses and ${\bf k}=(k_x,k_y,k_z)$. The energy and $\mu$ are measured relative to the bottom of the Q1D band. The lowest energy of the 3D band is negative $\varepsilon_{0} < 0$ and, to have a BCS-like condensate in the 3D band, we assume $|\varepsilon_{0}| \gg \mu$. Our study is focused on the superconducting state near the Lifshitz transition at $\mu=0$. The system is considered in the clean limit, where the role of impurities is neglected. In what follows we set $k_B=1$ for the Boltzmann constant.

Following Refs.~\cite{Suhl1959,Mos1959},  the mean-field Hamiltonian of the two-band superconductors is written as 
\begin{align}
{\cal H}= &\int {\rm d}^3{\bf r} \Big\{\sum\limits_{\nu=1,2}\,
\Bigl[\sum\limits_{\sigma}{\hat \psi}^{\dagger}_{\nu\sigma}({\bf r})\,T_\nu({\bf r}) \,{\hat
\psi}_{\nu\sigma}({\bf r})\notag\\
&+ \big({\hat \psi}^{\dagger}_{\nu\uparrow}({\bf r}){\hat \psi}^{\dagger}_{\nu\downarrow}({\bf r})\, \Delta_{\nu}({\bf r}) + {\rm h.c.}\big) \Bigr]+ \langle \vec\Delta, \check{g}^{-1}\vec\Delta \rangle \Big\}, \label{H}
\end{align}
where ${\hat\psi}^{\dagger}_{\nu\sigma}({\bf r})$ and ${\hat\psi}_{\nu\sigma}({\bf r})$ are the field operators for the carriers in band $\nu$, $T_{\nu}({\bf r})$ is the single-particle Hamiltonian with the single-particle energies given by Eq.~(\ref{disp12}), and $\Delta_{\nu}({\bf r})$ is the superconducting gap function for band $\nu$. We also use the vector notations $\vec\Delta = (\Delta_1,\Delta_2)$ with $\langle.,.\rangle$ the scalar product in the band vector space, and $\check{g}^{-1}$ is the inverse of the coupling matrix. The band-dependent superconducting gap functions satisfy the self-consistency condition given by the matrix gap equation
\begin{align}
\vec\Delta = \check{g} \vec{R},
\label{self}
\end{align}
where components of $\vec{R}$ are the anomalous Green functions $R_{\nu}=\langle\hat \psi_{\nu\uparrow}({\bf r}){\hat \psi}_{\nu \downarrow}({\bf r})\rangle$. 

The model based on Eqs.~(\ref{H})-(\ref{self}) is used to calculate the mean-field critical temperature $T_{c0}$ and then the fluctuation-shifted $T_c$. $T_{c0}$ is obtained by solving the linearized variant of the gap equation (\ref{self}). The fluctuations are investigated by using the expansion for the free energy functional for the two-band system with respect to the band superconducting gap functions, which essentially gives the two-band Ginzburg-Landau (GL) free energy functional. 

Assuming that $T_{c0}$ is known, one expands the r.h.s. of Eq.~(\ref{self}) with respect to $\Delta_\nu$. The lowest order terms of this expansion are given by~\cite{Geilikman1967, Asker2002, Zhit2004, Kogan2011, Shan2011, Vagov2012, Orlova2013} 
\begin{align}
R_{\nu}[\Delta_{\nu}] =&({\cal A}_{\nu} - a_{\nu})\Delta_{\nu} - b_{\nu}\Delta_{\nu}|\Delta_{\nu}|^2 \notag\\
&+\sum\limits_{i=x,y,z}{\cal K}^{(i)}_{\nu}\nabla_{i}^2\Delta_{\nu},
\label{R}
\end{align}
where the coefficients ${\cal A}_{\nu}$, $a_{\nu}$, $b_{\nu}$, and ${\cal K}^{(i)}_{\nu}$ are to be calculated using the microscopic model for each band, and external fields are assumed to be zero. 

For the 3D BCS-like band with the spherically symmetric Fermi surface one obtains the standard expressions 
\begin{align}
&{\cal A}_1=N_1 \ln \left( \frac{2 e^{\gamma} \hbar \omega_c}{\pi T_{c0}} 
\right), \, a_1=  -\tau N_1,\, b_1 = \frac{7\zeta(3)}{8\pi^2}\frac{N_1}{T^2_{c0}},\notag\\ 
&{\cal K}^{(x)}_1={\cal K}^{(y)}_1={\cal K}^{(z)}_1=\frac{\hbar^2v_1^2}{6} b_1,
\label{coeff1}
\end{align}
where $\tau=1-T/T_{c0}$, $\hbar\omega_c$ is the energy cutoff (assumed to be the same for both bands), $\gamma$ is the Euler constant, $\zeta(x)$ is the Riemann zeta function, the DOS of the 3D band at the Fermi energy is $N_1=m_1k_F/2\pi^2\hbar^2$ and the 3D band Fermi velocity $v_1=\hbar k_F/m_1$ is determined by the corresponding Fermi wavenumber $k_F =\sqrt{2m_1(\mu-\varepsilon_{0})}/\hbar$.   

For the Q1D band the expressions for the coefficients are given by the integrals to be evaluated numerically. At $|\mu| < \hbar\omega_c$~(near the Lifshitz transition) the coefficients can be written as
\begin{align}
&{\cal A}_2= N_2 \int\limits_{-\tilde{\mu}}^{1}\mbox{d} y\,  \frac{\tanh \big(  y / 2\tilde T_{c0} \big)}{y \sqrt{y+\tilde{\mu}}},  \notag\\
& a_2=-\tau\frac{N_2}{2\tilde T_{c0}} \int\limits_{-\tilde{\mu}}^{1}\mbox{d} y \,  \frac{\text{sech}^2  \big(  y / 2\tilde T_{c0} \big) }{\sqrt{y+\tilde{\mu}}}, \notag \\
& b_2 = \frac{N_2}{4 \hbar^2 \omega_c^2} \int\limits_{-\tilde{\mu}}^{1} \mbox{d} y\,  \frac{\text{sech}^2  \big(  y / 2\tilde T_{c0} \big) }{y^3 \sqrt{y+\tilde{\mu}}}  \left[ \sinh \Big( \frac{y}{\tilde T_{c0}} \Big)- \frac{y}{\tilde T_{c0}}\right] , \notag \\
& \mathcal{K}^{(z)}_2 = \hbar^2 v_2^2  \frac{N_2}{8\, \hbar^2 \omega_c^2} \int\limits_{-\tilde{\mu}}^{1} \mbox{d} y\,  \frac{\sqrt{y+\tilde{\mu}} }{y^3} \, \text{sech}^2  \big(  y / 2\tilde T_{c0} \big)  \notag \\
& \quad \quad \times  \left[ \sinh \left( \frac{y}{\tilde T_{c0}} \right)- \frac{y}{\tilde T_{c0}}\right] , \quad  {\cal K}^{(x,y)}_2=0,
\label{coeff2}
\end{align}
where we use the scaled quantities $\tilde{T}_{c0} = T_{c0}/\hbar \omega_c$ and $\tilde{\mu} = \mu /\hbar \omega_c$, and the effective band velocity $v_2$ is determined by the cutoff energy as $v_2 = \sqrt{2\hbar\omega_c/m_2}$~(independent of $\mu$). The effective DOS of the Q1D band is given by $N_2=\sigma_{xy}/4\pi \hbar v_2$, where the factor $\sigma_{xy}$ accounts for the contribution to DOS by the $x,y$ dimensions.  

The mean-field critical temperature $T_{c0}$ is obtained by solving the linearized gap equation which reads as [see Eqs.~(\ref{self}) and (\ref{R})]
\begin{align}
\check{L}\vec{\Delta} = 0,\quad \check{L}=\check{g}^{-1}
-\left(\begin{array}{cc} {\cal A}_1 & 0\\
0 & {\cal A}_2
\end{array}
\right). 
\label{lingap}
\end{align}
This is the matrix equation with the solution in the form 
\begin{align}
\vec{\Delta}=\psi({\bf r})\vec{\eta},
\label{lingap1}
\end{align}
where $\vec{\eta}$ is an eigenvector of $\check{ L}$ corresponding to its zero eigenvalue while $\psi({\bf r})$ is a coordinate dependent  GL order parameter of the system~\cite{Vagov2012, Orlova2013}.  A non-trivial solution to Eq. (\ref{lingap}) exists only when the determinant of $\check{ L}$ is zero, which gives the equation
\begin{equation}
(g_{22}-G {\cal A}_1)(g_{11}-G {\cal A}_2)-g^2_{12}=0,
\label{Tc0}
\end{equation}
with $G=g_{11}g_{22}-g^2_{12}$. Of the two solutions to Eq.~(\ref{Tc0}), one chooses the maximal $T_{c0}$. The corresponding eigenvector $\vec{\eta}$ can be adopted in the form
\begin{align}
\vec{\eta}=
\left(\begin{array}{c}
S\\
1
\end{array}\right), \; S=\frac{g_{11} - G {\cal A}_2}{g_{12}}.
\label{eta}
\end{align}
Notice that this choice is unique up to the normalization factor which is absorbed by $\psi ({\bf r})$.

The actual critical temperature $T_c$ is lower than its mean field value $T_{c0}$ due to fluctuations~\cite{Varlamov}. The fluctuation-induced correction to $T_{c0}$ is obtained by using the standard Gibbs distribution $\exp(-F/T)$, where the free energy functional writes as (see e.g. \cite{Asker2002,Zhit2004})
\begin{equation}
F = \int d^3{\bf r}\Big[\sum_{\nu=1,2}   f_\nu  +  \langle\vec\Delta, \check{L}
\vec\Delta\rangle\Big],   
\label{free1}
\end{equation}
with
\begin{equation}
f_\nu = a_\nu \left|\Delta_{\nu} \right|^2  + \frac{b_{\nu}}{2} \left|
\Delta_{\nu} \right|^4 +\sum\limits_{i=x,y,z} {\cal K}^{(i)}_{\nu}  \left|\nabla_i\Delta_{\nu} 
\right|^2.
\label{free2}
\end{equation}
The stationary condition for the functional given by Eqs.~(\ref{free1}) and (\ref{free2}) yields the gap equation (\ref{self}).

The calculations of the fluctuation corrections are simplified by representing $\vec{\Delta}$ as a linear combination of the vectors $\vec{\eta}$ and  $\vec{\xi}=(1,-S)^T$~[one can see that $\langle\vec{\eta},\vec{\xi}\rangle=0$] 
\begin{align}
\vec{\Delta} ({\bf r}) = \psi ({\bf r})\vec{\eta} +\varphi({\bf r}) \vec{\xi},
\label{exp}
\end{align}
where $\varphi({\bf r})$ is the second fluctuation mode. The free energy functional is then expressed in terms of $\psi$ and $\varphi$ as
\begin{align}
F  = \int d^3{\bf r}(f_\psi  + f_\varphi + f_{\psi \varphi}),
\label{free3}
\end{align}
where $f_\psi$ and $f_\varphi$ have the same structure given by Eq.~(\ref{free2}), where $\Delta_\nu$ is replaced by $\psi({\bf r})$ and $\varphi({\bf r})$, respectively, and the set of the coefficients $\{\alpha_\nu, b_\nu, {\cal K}_{\nu}\}$ is changed to $\{\alpha_{\psi}, b_{\psi}, {\cal K}_{\psi} \}$ and $\{ \alpha_{\varphi}, b_{\varphi}, {\cal K}_{\varphi} \}$. In addition, $f_{\psi \varphi}$ in Eq.~(\ref{free3}) represents the coupling between the two modes $\psi({\bf r})$ and $\varphi({\bf r})$. The coefficients in $f_{\psi}$ one obtained as
\begin{align}
&a_{\psi}=S^2  a_1 + a_2,\; b_{\psi} = S ^4 b_1 + b_2,\notag\\
&{\cal K}^{(i)}_{\psi} = S^2{\cal K}^{(i)}_1 + {\cal K}^{(i)}_2,
\label{coeffpsi}
\end{align}
whereas the coefficients in $f_{\varphi}$ are given by  
\begin{align}
&a_{\varphi} = a^{(0)}_{\varphi}+ a_1 +S^2 a_2, \; b_{\varphi} = b_1 + S^4b_2,\notag\\
&{\cal K}^{(i)}_{\varphi} = {\cal K}^{(i)}_1 + S^2{\cal K}^{(i)}_2,
\label{coeffphi}
\end{align}
with
\begin{equation}
\alpha^{(0)}_{\varphi}=\frac{(1+S^2)^2}{SGg_{12}}.
\label{coeffphi1}
\end{equation}
Here $\alpha^{(0)}_{\varphi} \neq 0$ since $S$ is real. This means that only $f_\psi$ represents the critical fluctuations in the vicinity of the superconducting transition because $\alpha_\psi \to 0$ in the limit $T \to T_{c0}$. The contribution $f_{\varphi}$ describes non-critical fluctuations and can be safely omitted~\cite{Salasnich2019}. Thus, the fluctuations are determined by the GL functional $f_\psi$, with the single component order parameter $\psi({\bf r})$. Due to the presence of the Q1D band, this functional is anisotropic with ${\cal K}^{(x,y)}_{\psi} \neq {\cal K}^{(z)}_{\psi}$.

\begin{figure*}[t]
\centering
\hspace{-0.6cm}\includegraphics[width=1.0\linewidth]{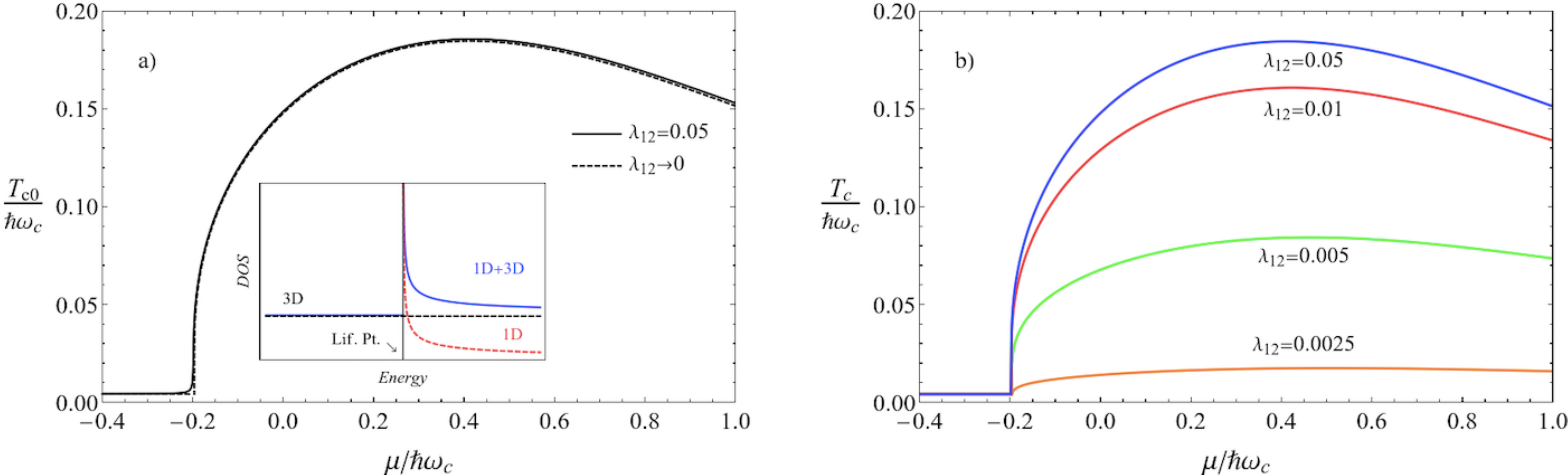}
\caption{a) The mean-field critical temperature $T_{c0}$ versus the chemical potential $\mu$, calculated for $\lambda_{12}=0.05$ and $\lambda_{12} \to 0$; the insert demonstrates the energy dependent DOSs with the 
van Hove singularity of the Q1D DOS at the Lifshitz point $\mu=0$. b) The fluctuation-shifted critical temperatures $T_c$ as a function of $\mu$, calculated for selected values of the dimensionless pair-exchange coupling constant $\lambda_{12}=0.001,0.0025,0.005,0.05$ at the Ginzburg number $Gi_{3D}=10^{-10}$ of the uncoupled 3D band.}
\label{fig1}
\end{figure*}

With this simplification, we can apply the known results for the fluctuation-driven shift of the critical temperature in the single-component GL theory. Using the renormalization group approach, one obtains~\cite{Varlamov} that the actual 3D critical temperature is related to the mean-field one by  
\begin{equation}
\frac{T_{c0}-T_c}{T_c}=\frac{8}{\pi}\sqrt{Gi},
\label{shiftTc}
\end{equation}
where $Gi$ is the Ginzburg number (Ginzburg-Levanyuk parameter). For the 3D anisotropic system it reads~\cite{Varlamov}
\begin{equation}
Gi=\frac{1}{32\pi^2}\, \frac{T_{c0}b^2_{\psi}}{a^{\prime}_{\psi}{\cal K}^{(x)}_{\psi}{\cal K}^{(y)}_{\psi}{\cal K}^{(z)}_{\psi}},
\label{Gi}
\end{equation}
with $a^{\prime}_\psi=da_{\psi}/dT$. Using Eq.~(\ref{coeffpsi}), the above expression can be rearranged as
\begin{equation}
Gi = Gi_{3D} \frac{ \left(b_2/b_1+S^4\right)^2}{\big(a_2/a_1+S^2\big)\Big({\cal K}^{(z)}_2/{\cal K}^{(z)}_1+S^2\Big)S^4},
\label{Gi1}
\end{equation}
where $Gi_{3D}$ is the Ginzburg number of the uncoupled (standalone) 3D band, given by Eq.~(\ref{Gi}) with the substitution $\{ a_{\psi}, b_{\psi},  {\cal K}^{(i)}_{\psi} \} \to \{a_1, b_1, {\cal K}^{(i)}_{1} \}$.
 
Using the obtained expressions, we can now calculate both the mean-field $T_{c0}$ and fluctuation-shifted $T_c$ critical temperatures. Essential parameters of the model are the three coupling constants $g_{\nu \nu^\prime}$ and the band DOSs $N_\nu$, while the cutoff $\hbar\omega_c$ sets the energy scale. It is convenient to introduce the dimensionless coupling constants $\lambda_{\nu \nu^\prime} =g_{\nu \nu^\prime} \sqrt{N_\nu N_{\nu^\prime}}$. The parameter $S$, which controls Eqs.~(\ref{shiftTc})-(\ref{Gi1}) and also $T_{c0}$, depends on $\lambda_{11},\lambda_{22},\lambda_{12}$ as well as on the ratio $N_2/N_1$. In the calculations we assume $\lambda_{22}=0.2$ and $\lambda_{11}=0.18$, which is in the range of typical values of the dimensionless couplings in conventional weak-coupling superconductors~\cite{Fetter}. We also take $N_2/N_1 = 1$ for simplicity. Finally, we need also to specify $N_1$ which defines $Gi_{3D}$. We follow a different path and use an estimate $Gi_{3D}=10^{-10}$ by taking into account that the Ginzburg number of most 3D superconductors is in the range $10^{-6 \div 16}$~\cite{Ketterson}, being $Gi_{3D}\approx (T_{c1}/E_F)^4$, with $T_{c1}$ the critical temperature of the standalone band $1$ and $E_F=\hbar^2k^2_F/2m_1$~(for the stable 3D condensate $T_{c1}\ll E_F$). Notice, that our results are not sensitive to a particular choice of the microscopic parameters unless the dimensionless intraband coupling of the 3D band is significantly larger than that of the Q1D band and the two-band system approaches a routine 3D superconductor. 

Numerical results for $T_{c0}$ and $T_c$ versus the chemical potential $\mu$, calculated for several values of the dimensionless pair-exchange coupling $\lambda_{12}$, are shown in Fig.~\ref{fig1}. One sees from Fig.~\ref{fig1} a) that when $\mu$ is sufficiently below zero, the Q1D band does not contribute and $T_{c0}$ is determined by the uncoupled band $1$. In the vicinity of the Lifshitz transition at $\mu = 0$, $T_{c0}$ rises sharply by a factor of $40$. At larger $\mu$, $T_{c0}$ decreases, approaching the critical temperature of the decouple 3D band again. As is well known, the reason for this sharp rise is the increase in the energy-dependent Q1D DOS that has the van Hove singularity at the band edge, as illustrated in the inset in Fig.~\ref{fig1} a). It is remarkable that $T_{c0}$ is almost insensitive to the pair-exchange coupling as long as $\lambda_{12} \ll \lambda_{11}$. Consequently, in the vicinity of the Lifshitz transition, the superconducting properties of the two-band system on the mean field level are fully determined by the Q1D band.

In contrast, the fluctuation-induced shift of the critical temperature strongly depends on the pair-exchange coupling. In the limit $\lambda_{12} \to 0$ the fluctuations suppress the superconductivity. However, this suppression ceases rapidly with increasing the pair-exchange coupling. Figure~\ref{fig1} b) demonstrates that the presence of even a vanishingly small coupling, $\lambda_{12} \ll \lambda_{\nu\nu}$, is sufficient to quench the fluctuations and to eliminate the shift. In particular, $T_{c}$ approaches $T_{c0}$ already at $\lambda_{12} \simeq 0.01$ and at $\lambda_{12} \simeq 0.05$ the two critical temperatures are practically indistinguishable. 

Concluding, our calculations demonstrate that coupling to a stable 3D condensate ``screens" out the severe thermal fluctuations of the Q1D superconducting condensate. This coupling gives rise to a single critical mode that controls the thermal fluctuations of the condensate gap functions $\Delta_1({\bf r})$ and $\Delta_2({\bf r})$. In other words ``light" excitations of the Q1D condensate are always accompanied by ``heavy" excitations of the stable 3D condensate.  Therefore, such a two-band system becomes a robust mean-field superconductor. 

The superconductivity enhancement, based on the interaction between a Q1D condensate near the Lifshitz point and a BCS-like condensate, is general and does not depend on the model details. Thus, it opens a possibility for a significant amplification (up to orders of magnitude) of the critical temperature by tuning the Lifshitz transition. Notice that in addition to the chemical engineering, Lifshitz transitions can be tuned by an appropriate doping of multiband superconducting compounds, as e.g. reported in \cite{Wu2019}.

Finally, we note that although our results are obtained for the model with the $s$-wave pairing, one can expect that the fluctuations screening mechanism, based on the coupling of multiple condensates, applies also to materials with the $d$-wave symmetry and even to the case of the triplet pairing, having the multicomponent order parameter. In this regard we note that first theoretical calculations of the possible pairing symmetry in Q1D multiband superconductors ${\rm A}_2{\rm Cr}_3{\rm As}_3$~(with ${\rm A}={\rm K},\,{\rm Rb},\,{\rm Sc}$) are in favour of the triplet pairing~\cite{Wu2015}. A detailed analysis of the fluctuations for the triplet pairing requires a separate investigation.

A.V. acknowledges support of the CAPES/Print Grant, process No. 88887.333666/2019-00 (Brazil), and from the Russian Science Foundation Project 18-12-00429. T.T.S. acknowledges financial support and hospitality by the Universities of Camerino and Padova.

\end{document}